\documentclass[12pt]{iopart}

\usepackage{iopams}

\usepackage{wasysym}  
\usepackage{marvosym}  
\usepackage{amssymb}  
\usepackage{pifont}  
\usepackage{textcomp}  

\usepackage{graphicx}

\begin{document}

\title[Classification of Ge hut clusters in the arrays  formed on the Si(001) surface\dots]{Classification of Ge hut clusters in the arrays  formed by molecular beam epitaxy at low temperatures  on the Si(001) surface }

\author{Larisa V Arapkina and Vladimir A Yuryev\footnote[1]{http://www.gpi.ru/eng/staff\_s.php?eng=1\&id=125}}

\

\address{A.\,M.\,Prokhorov General Physics Institute of the Russian Academy of Sciences,\\ 38 Vavilov Street, Moscow, 119991, Russia}
\ead{arapkina@kapella.gpi.ru}

\begin{abstract}
Morphological investigations and classification of Ge hut clusters forming the arrays of quantum dots on the Si(001) surface at low temperatures in the process of the ultrahigh vacuum molecular beam epitaxy have been carried out using {\it in situ} scanning tunnelling microscopy. Two main species of Ge hut clusters composing the arrays---pyramidal and wedge-shaped ones---have been found to have different atomic structures. The inference is made  that shape transitions between pyramids and wedges are impossible. The nucleation probabilities of pyramids and wedges equal $1/2$ at the initial stage of the array formation. 
The wedges become the dominating species as the amount of the deposited germanium is increased. A fraction and a density of the pyramids in the arrays are rapidly decreased  with the growth of Ge coverage. 

The derivative types of the clusters---obelisks (or truncated wedges) and accreted wedges---have been revealed and investigated for the first time, they have been found to start dominating  at high Ge coverages. The obelisks originate from the wedges as a result of their height limitation and further growth of trapezoid facets. The apexes of the obelisks are formed by  sets of the parallel (001) ridges. 

The uniformity of the cluster arrays have been evidenced to be controlled by the length distribution of the wedge-like clusters. At low growth temperatures ($360^\circ$C) nucleation of new clusters is observed during the array growth at all values of Ge coverage except for a particular point at which the arrays are  more uniform than at higher or lower coverages. At higher temperatures ($530^\circ$C)  cluster nucleation has not been observed after the initial stage of the array formation.

\end{abstract}


\maketitle

\section{\label{sec:intro}Introduction}

\subsection{\label{sec:general}Problem definition}

The development of processes of the controllable formation of germanium quantum dot (QD) arrays on the silicon surface as well as multilayer Ge/Si epitaxial heterostructures on their basis is a subject of significant and permanently increasing efforts for a number of years \cite{Report_01-303,Pchel_Review,Brunner,Wang-properties} primarily due to their potential applications in prospective devices of microelectronics and integrated microphotonics compatible with the monolithic silicon VLSI technology.  Both high density of the germanium nanoclusters ($> 10^{11}$~cm$^{-2}$) and high uniformity of the cluster shapes and sizes (dispersion $<$ 10\,\%)   in the arrays are required for many practically important application of such structures \cite{Pchel_Review,Wang-properties,Smagina,Wang-Cha,QDIP-Wang,QDIP-Wang1, WG-near_IR,Photonic_crystal,QDFET}. 

The main technique of formation of the germanium nanoclusters on the silicon surface is the molecular beam epitaxy (MBE) \cite{Pchel_Review,Brunner}. A high density of the self-assembled clusters can be obtained in the MBE process of the Ge/Si(001) structure formation when depositing germanium on the silicon substrate heated to a moderate temperature ($\lesssim 550^\circ$C).\footnote{ Remark also that lowering of the array formation temperature down to the values of $\lesssim 450^\circ$C is required for the compatibility of the Ge/Si(001) heterostructure formation process with  the CMOS device fabrication cycle \cite{Report_01-303}. It is another reason to decrease the temperature of all treatments starting from the Si surface preparation. } In this case the lower  the temperature of the silicon substrate in the process of the germanium deposition the higher  the density of the clusters is at the permanent quantity of the deposited germanium \cite{Report_01-303,Yakimov,Jin}. For example the density of the germanium clusters in the array reached  $6\times 10^{11}$~cm$^{-2}$ at the substrate temperature during the deposition $T_{\rm gr} = 360^\circ$C and the effective thickness of the deposited germanium layer\footnote{I.e. the Ge coverage or more accurately the thickness of the Ge film measured by the graduated in advance film thickness monitor with the quartz sensor installed in the MBE chamber.} $h_{\rm Ge} = 8~{\rm\AA}$ whereas the cluster density of only about $2\times 10^{11}$~cm$^{-2}$ was obtained at  $T_{\rm gr} = 530^\circ$C and the same value of $h_{\rm Ge}$  \cite{Report_01-303}. 

There are also different ways to increase the cluster density in the arrays. Thus, the authors of Ref.~\cite{Smagina}  succeeded to reach the cluster density of about $9\times 10^{11}$\,cm$^{-2}$ in the array using the pulsed irradiation of the substrate by a low-energy Ge$^+$ ion beam during the MBE growth of the Ge/Si(001) heterostructure at  $T_{\rm gr} = 570^\circ$C.

\begin{figure}[h]
\centering
\includegraphics[scale=1]{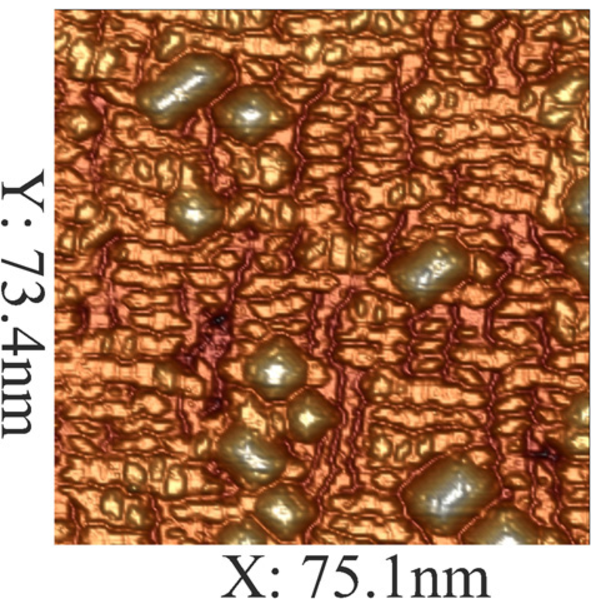}(a)
\includegraphics[scale=1]{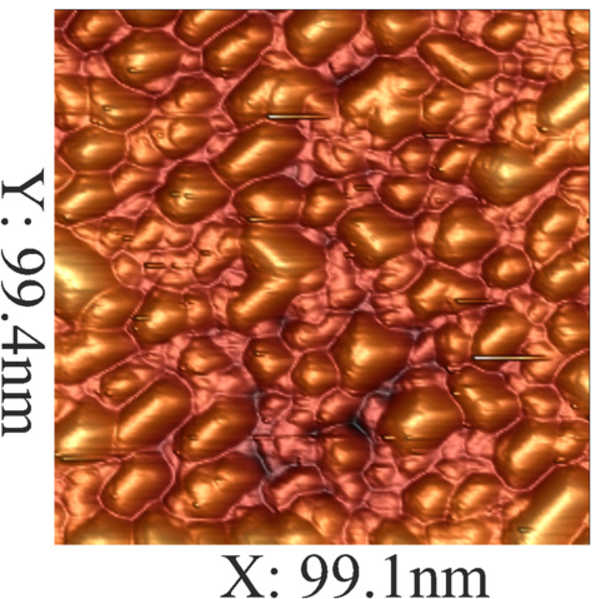}(b)\\
\includegraphics[scale=1]{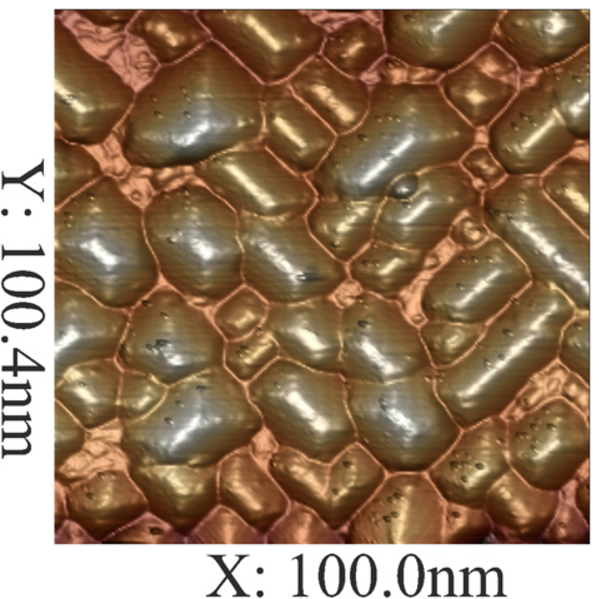}(c)
\includegraphics[scale=1]{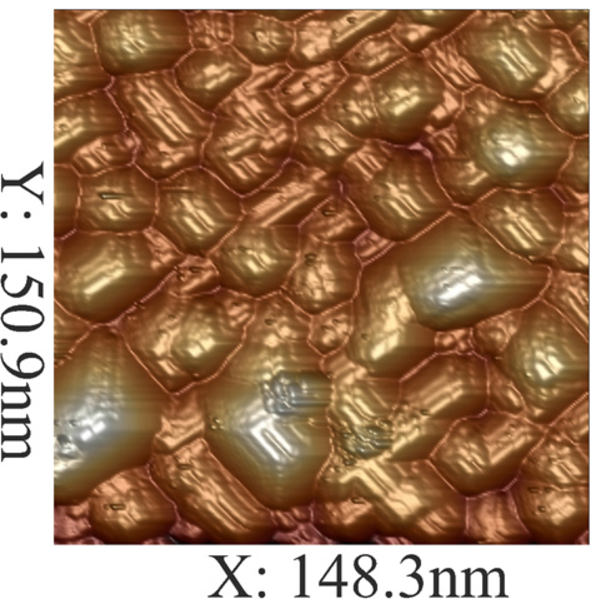}(d)
\caption{\label{fig:array_360C}{\it In situ} STM micrographs of the Ge hut cluster arrays formed on the Si(001) surface in the process of molecular beam epitaxy at  the substrate temperature  $T_{\rm gr} = 360^\circ$C and different effective thicknesses of the deposited Ge layer $(h_{\rm Ge})$, the values of $h_{\rm Ge}$ are (a) 6~\r{A}; (b) 8~\r{A}; (c) 10~\r{A}; (d) 14~\r{A}. Diagonals of the images are parallel to the  $\langle$1\,0\,0$\rangle$ axes.}
\end{figure}  

\begin{figure}[h]
\centering
\includegraphics[scale=1]{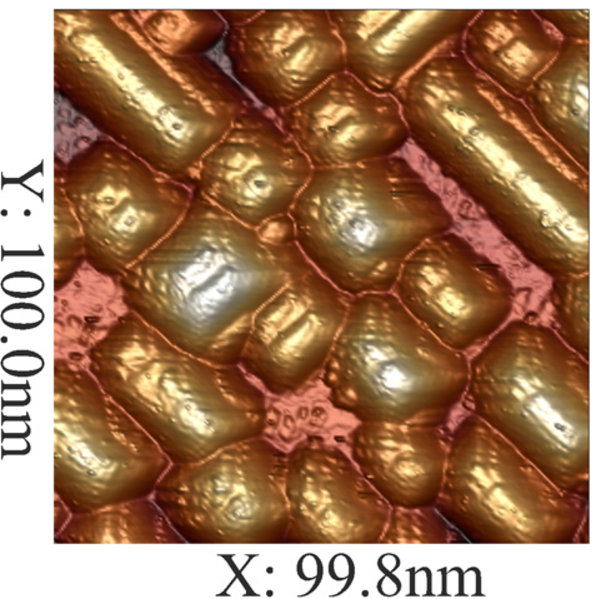}(a)
\includegraphics[scale=1]{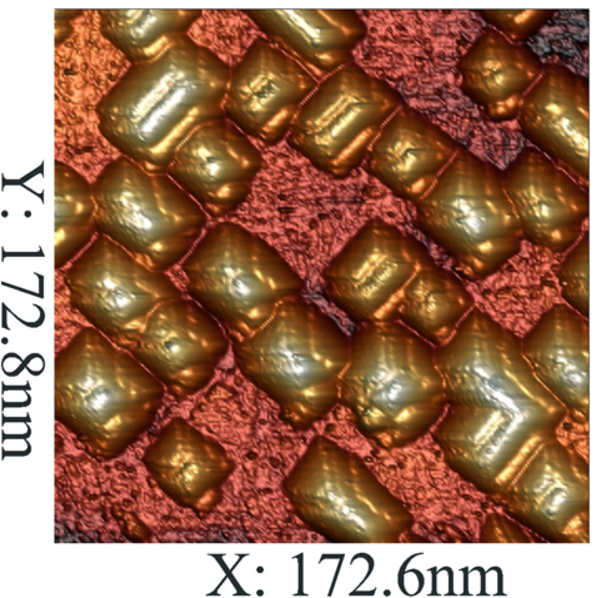}(b)\\
\includegraphics[scale=1]{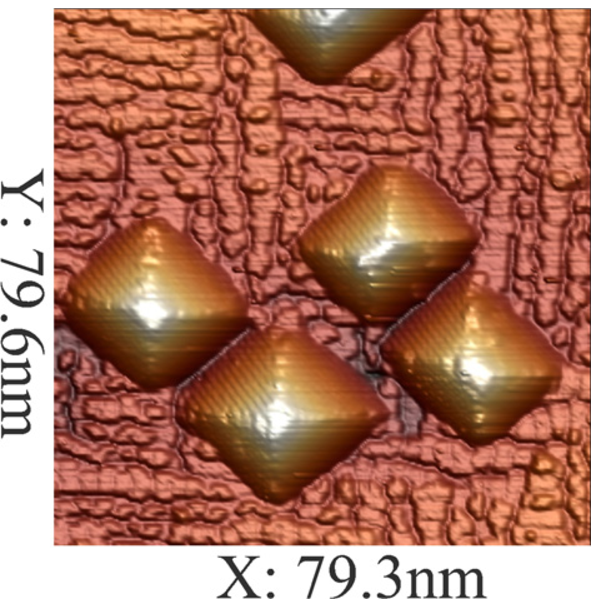}(c)
\caption{\label{fig:array_530C}The same as in Fig.~\ref{fig:array_360C} but for $T_{\rm gr} = 530^\circ$C,  $h_{\rm Ge}$: (a) 8~\r{A}; (b) 10~\r{A}; (c) 11~\r{A}. }
\end{figure}

Obtaining of the arrays of the densely packed Ge QDs on the Si(001) surface is an important task but the problem of formation of  uniform arrays of the Ge clusters  is even more challenging one. The process of Ge/Si(001) heterostructure formation with the Ge QD dense arrays and predetermined electrophysical and photoelectric parameters cannot be developed until both of these tasks are  solved. The uniformity of the cluster sizes and shapes in the arrays determines not only the widths of the energy spectra of the charge carrier bound states in the QD arrays  \cite{Smagina} but in a number of cases the optical and electrical properties of both the arrays themselves and the  device structures produced on their basis \cite{Electrolumin}. To find an approach to the improvement of the Ge QD array uniformity on the Si(001) surface it is necessary to carry out the morphological investigation of the clusters constituting the arrays and first of all classify them.

At present, two genera of the self-assembled Ge clusters formed on the Si(001) surface are marked out---huts and domes. The former are smaller and faceted by $\{105\}$ planes while the later are much larger and have more sophisticated faceting \cite{Wang-Cha,Jin,Ge_Shapes, Mo, Vailionis, Ross}. Our investigations of the densely packed Ge nanocluster arrays on the Si(001) surface \cite{Report_01-303} have shown that the composition of an ensemble of the hut clusters is by no means homogeneous---there are several species of the hut clusters different by their geometrical shapes as well as their behaviour in the process of the array formation.\footnote{Here we do not consider defects of the arrays. Other article will be devoted to their investigation. Now the available information about the morphology, structure and density of the array defects as well as their effect on the array parameters can be found in our reports \cite{Report_01-303} or \cite{defects_ICDS-25}.} 
Examples of the Ge cluster arrays formed on the Si(001) surface at different $T_{\rm gr}$ and $h_{\rm Ge}$ are shown in Figs.~\ref{fig:array_360C} and \ref{fig:array_530C}  which obviously demonstrate that the arrays grown at low temperature always consist of a set of morphologically different hut clusters. Some of them have often been  discussed in the literature since the classical letter by Mo {\it et al.} \cite{Mo, Vailionis, Tersoff_Tromp,  Nucleation, Goldfarb_JVST-A, Island_growth, Kastner,  Goldfarb_2005, Goldfarb_2006,Jesson-MRS} and some have not.

This paper is devoted to the study of morphological differences of the Ge hut clusters formed in the process of molecular beam epitaxy at low substrate temperatures on the Si(001) surface and their classification derived from the revealed species differences.

Writing about the classification of the hut clusters we have not got a goal to simply introduce a new terminology, as it might seem, although the latter is also proposed.\footnote{Since the pioneering paper by Mo {\it et al.} \cite{Mo}, a good few of descriptive and often confusing terms have been used in the literature to designate two known types of the hut clusters. In this paper, we introduce a new strict stereometrical terminology to  emphasize the structural difference between the cluster species and avoid muddle in the future. 
We shall name each species of the clusters in accordance with the denominations of the geometrical bodies which  most accurately describe  the shapes of the clusters.} The aim of classification is to sort out the hut clusters in accordance with their structural peculiarities which are much more important attributes  than geometrical shapes. The difference of geometrical shapes does not necessarily imply the difference of atomic structures. The latter may be identical. Hence, if so,  it may be assumed that one type of clusters originates from another, as it is usually accepted ``by default'' (pastulated) for pyramids and ``elongated'' hut clusters in the literature \cite{Kastner,  Goldfarb_2005, Goldfarb_2006}. On the contrary, if the atomic structure of two species of clusters is different the shape transitions seem to be very unlike because of a probably high potential barrier to be overcome to change the cluster atomic configuration and symmetry.

\subsection{\label{sec:excursus}Formation of Ge hut clusters. A brief excursion into the historical domain}

It is well known \cite{Chem_Rev} that in equilibrium Ge layers grow on the Si(001) surface following the Stranski-Krastanow  mode \cite{Stranski-Krastanow,Growth_modes}. It means that initially Ge grows layer by layer until reaches a thickness of a few monolayers, then nucleation of the three-dimentional islands begins \cite{Pchel_Review}. 

 Because of lower surface energy, Ge when growing wets the Si surface (this enables tracing an etymology of the term ``wetting layer''). Despite the mismatch of Si and Ge lattices, which is $\sim 4.2$\,\%, Ge atoms deposited on the Si(001) surface keep the correspondence with the Si atoms for several monolayers. The Ge layer surface is dimerized and  $(2\times 1)$ reconstructed \cite{Chem_Rev}. Due to dimer buckling it comprises a mixture of the $c(4\times 2)$ and $p(2\times 2)$ structures \cite{Iwawaki_dimers,Iwawaki_initial} which are manifested in STM images as characteristic antiphased and cophased zigzags. 
As Ge atoms arrive onto the surface, the compression along the dimer rows is relieved by arising dimer vacancies. Further ordering of the dimer vacancies in a nearly regular array of parallel trenches (or formation of the so called $(2\times n)$ reconstruction \cite{Chem_Rev}), which are perpendicular to the dimer rows,  goes on to reduce the surface strain energy beyond the Ge coverage of 0.8 monolayer (thickness of 1\,ML $\approx 1.4$\,\r{A}) \cite{Wu}. And at last, the formation of the dimer-row vacancies and a grating  of the quasiperiodic $(M \times N)$ patched structure \cite{Iwawaki_initial,Wu,Voigt,Wetting} exhausts the ability of the dimer vacancies to accommodate the wetting layer to the strain increasing with the growth of quantity of the deposited Ge. When the Ge coverage exceeds 3\,ML, the three-dimentional nanoislands (or Ge nanoclusters) start to nucleate on the surface. They are free of dislocations in the Si/Ge interface (coherent with the Si substrate), faceted and have base sides  alined with  two orthogonal $\langle$1\,0\,0$\rangle$ axes.  At moderate growth temperatures the composition of the cluster arrays is bimodal: A part of clusters have shapes of regular square-based pyramids while others have rectangles in their bases \cite{Mo,Iwawaki_SSL}. Due to the shapes resembling huts both square-based and rectangular-based clusters are usually referred to as ``hut'' clusters. The hut clusters (coherent islands) were theoretically shown to be (under some conditions) more energetically stable  than the strained films or dislocated islands \cite{Tersoff_Tromp,Tersoff_LeGoues}. Their appearance was also found to be kinetically favourable compared to the nucleation of dislocations \cite{Tersoff_LeGoues}.

The first STM observation of hut clusters was reported by Mo {\it et al.} in 1990  \cite{Mo} (the term ``hut'' was introduced in that article). That letter presented an experimental investigation of a newly discovered metastable phase of Ge clusters which arose  in the process of Stranski-Krastanow  growth   to relieve the increasing  stress in the wetting layer when Ge was deposited  by MBE at moderate temperature on Si(001). As distinct from  the macroscopic clusters, hut clusters were found to have $\{105\}$ faceting on all sides. The first model of the cluster $\{105\}$ faces according to which these facets consist of (001) terraces composed by {\it pairs of dimers} of the Ge(001)-$(2\times 1)$ reconstructed surface   was suggested in that communication (now this model is usually referred to as the PD model \cite{Iwawaki_SSL,Fujikawa_ASS,Facet-105}).  Huts were found to have ``predominantly {\it a prism shape} (with canted ends), in some cases four-sided pyramids,  with the same atomic structure 
on all four facets''. The observed length of the huts was really huge (up to 1000 \r{A}) while their widths did not exceed 200~\r{A}  (the aspect ratio often reached 10).
The question was put for the first time in that article what caused the elongation and specific base orientation of huts. Unfortunately,  no convincing answer to this question has been proposed thus far.

It should be noted that since the letter by Mo {\it et al.}, pyramidal and ``elongated'' clusters have alway been considered in the literature as structurally identical ones different only by their base aspect ratio \cite{Nucleation,Goldfarb_JVST-A,Island_growth,Kastner,Goldfarb_2005,Goldfarb_2006,Jesson-MRS}. The only argument for this assumption---the identity of faceting---does not seem to us to be very solid. Stress relaxation via formation of the $\{105\}$ faceted structures, such as islands or pits \cite{Nucleation}, appears to be energetically favourable and the structures are more stable compared to ones with different faceting \cite{Facet-105}. This means that faceting itself cannot be considered as the only sign of belonging to some specific group of morphologically identical clusters. 
Total atomic structure of clusters is defined by both the structure of their facets and the configuration of their apexes. If the latter is different, the clusters should be regarded as members of different species. 

The sound counterargument to the assumption about the cluster identity, which is usually disregarded, was given, e.g., in Ref.~\cite{Stability-Anealing}: ``Elongated'' clusters completely disappeared from arrays after annealing at $550^\circ$C for 600\,s whereas pyramids and domes remained. Moreover, a commonly adopted pathway of the dome cluster formation is as follows: some ``prepyramid'' (we have never observed such formations in our low-temperature MBE experiments)\,$\rightarrow$\,pyramid\,$\rightarrow$\,dome \cite{Vailionis, Pyramid_to_dome}. ``Elongated'' huts have never been met on this pathway. 
This led us to suggestion that ``elongated'' huts\footnote{Hereinafter we shall, as a rule, refer to them as wedges, wedge-like or wedge-shaped clusters.} differ from pyramids not only by shapes but mainly by their atomic structure and, that is more important, by the genesis and the role in the array development.

Nevertheless, the question about the mechanism of hut elongation was asked and required answering. Before long, the simplest and at first glance the most probable scenario  was proposed and immediately adopted by the community. According to this scenario wedge arise by elongation of a pyramid due to the growth of one of its \{105\} facets \cite{Nucleation,Goldfarb_JVST-A,Island_growth,Kastner,Goldfarb_2005,Goldfarb_2006,Jesson-MRS}. 
This hypothesis would explain everything unless the necessity to explain why the symmetry of the pyramid is violated. Due to its shape the square-based regular pyramid seems to be stable enough, at least unless some exterior anisotropic agent affects it removing the degeneracy of its facets. Otherwise it is unclear why one facet gathers arriving Ge atoms to the prejudice of the rest three (or at least two) equivalent planes. 

The next often observed miracle is the formation of two closely neighbouring clusters separated by only a few nanometers, one of which is a pyramid while another is a wedge.  This means that the acting agent violating the system symmetry is localized within a few nanometers around the growing wedge and does not affect the adjacent growing pyramid. At high enough coverages such clusters often start to coalesce---this in turn means that the agent previously violated the symmetry of one cluster of the pair has reached the growing pyramid but now does not affect it! 

It may be, however, that the symmetry is violated only once at the beginning of the pyramid elongation, then agent stops acting and the cluster grows on triangular facets faster than on trapezoid ones. It should be concluded  now that, as distinct from the case of annealing \cite{Stability-Anealing}, pyramids are much less stable than wedges in the process of Ge deposition, perhaps due to supersaturation by Ge atoms on the surface. This conclusion seems to be in agreement with experimental data, although the nature of the fluctuating agent stays unclear.

Goldfarb {\it et al.} proposed that an asymmetry of the stress field due to the presence of \{105\} faceted pits may result in the cluster anisotropic lateral growth because of  so-called equilibrium-driven elongation \cite{Goldfarb_2006}. They found that in thick hydrogen rich  Ge wetting layer (between 7.7 and 8.3~ML), which is formed in the process of gas-source-molecular-beam-epitaxy (GS-MBE)\footnote[1]{GS-MBE much more resembles chemical vapour deposition (CVD) than the solid source ultrahigh vacuum (UHV) MBE in which an atomic beam of Ge supplies the growing layer with Ge atoms rather than a flux of GeH$_4$ or Ge$_2$H$_6$.}  growth of Ge on Si(001) from GeH$_4$,  the strain at $690^\circ$C is relieved   by formation of \{105\} faceted pits rather than islands \cite{Nucleation}, that is in agreement with the earlier conclusion by Tersoff and LeGoues \cite{Tersoff_LeGoues} according to which pits always have a lower energy than islands of the same shape and equal size. (Goldfarb {\it et al.} observed also that at $620^\circ$C---and hence thinner wetting layer---only hut clusters arose \cite{Nucleation}.) Further, with the increase of Ge coverage, when the capability of pits to release the stress  is exhausted hut clusters nucleate in the vicinity of pits \cite{Nucleation,Goldfarb_JVST-A,Goldfarb_2005}. Acknowledging a model by Jesson {\it et al.}, which explains instability of the hut cluster shapes (read ``elongation'') by nucleation and growth on the facets \cite{Jesson-MRS}, as more common, the authors of Ref.~\cite{Nucleation} illustrate the elongation process by an example of cluster coalescence. We agree with Goldfarb {\it et al.} that such event sometimes happens and will present below a picturesque  evidence of it. However, this explanation of the cluster elongation phenomenon of course by no means can be accepted as universal.\footnote{Note, by the way, that Goldfarb with co-authors were first who reported the STM observation of incomplete trapezoid facets of the wedges in Ref.~\cite{Goldfarb_JVST-A}. Until now, incomplete triangular facets have not been observed, however.} 

As mentioned above,  investigations by Goldfarb with co-authors \cite{Goldfarb_2006} eventually gave a weighty argument in support of the so-called equilibrium-driven elongation or, in other words, elongation governed by energy minimization. They considered the cluster evolution model by Tersoff and Tromp developed for conditions when an isolated faceted strained Stranski-Krastanow  island formed on a wetting layer grow in height slower than in lateral direction \cite{Tersoff_Tromp}. It may be shown in this case by minimizing the total energy per unit volume that there is a critical size of a pyramidal cluster ($\alpha_0$) and the cluster grows isotropically up to some point ($e\alpha_0$), then anisotropic elongation in one direction starts. The authors of Ref.~\cite{Goldfarb_2006} indicate the discrepancy of the estimates made on the basis of the model by Tersoff and Tromp \cite{Tersoff_Tromp} with experimental observations. The model predicts $e\alpha_0 \approx 100$\,nm ($\alpha_0 \approx 37$\,nm) whereas in opinion of the authors of Ref.~\cite{Goldfarb_2006}  anisotropic  elongation of huts starts at the nucleus size of $\sim$\,5--8\,nm \cite{Nucleation}, if at all.

Another model of the equilibrium driven growth was proposed by Li, Liu and Lagally for a two-dimentional rectangular island \cite{Goldfarb_2006,LLL}. The model is as follows: Let dimensions of the island  sides be $s$ and $l$. Till its size is less than some critical value, the total energy is minimum for $s=l$. For a greater island its square shape becomes instable because of the strain: The island begins elongation in one of two degenerate orthogonal directions until the total energy reaches minimum at some value of $\arctan(s/l)=\pi/4 \pm \Delta$. To explain the anisotropic elongation  the authors of Ref.~\cite{LLL}  have to introduce {\it anisotropy of edge energies} to violate the symmetry of a square. According to this model   island grows along the direction  of the lower edge  free energy in which both strain and edge energies are minimum. The advantage of this construct is that for strong anisotropy the elongation occurs at any size of the island. Goldfarb and co-authors extend this model to the case of a 3D faceted island with slowly growing height \cite{Goldfarb_2006}. According to them when island elongates its perimeter grows faster than the area of the strained  base in such a way resulting in more effective relaxation. Nevertheless, an origin of anisotropy stays an issue in this model. 

As mentioned above an explanation was however proposed in the same letter \cite{Goldfarb_2006}. It was observed that huts interact with parent pits as well as with different adjacent ones, their lengthwise growth often starts at one pit and ends at the other. Sometimes the elongation is finished when a hut grows along the pit boundary and reaches its corner. 
The effect of the hut and pit interaction was analyzed by finite element technique and found to cause hut elongation. The pit dimensions  used in calculations were $10\times 10$\,nm and the wetting layer was as thick as 2\,nm. 
The inference was made in the article that the following hut evolution scheme takes place: as soon as a stable critical nucleus  is formed \cite{Goldfarb_2005} it grows in the energetically favorable direction along its mutual boundary with the parent pit until reaches the pit corner or attains an equilibrium $s/l$ ratio. This finishes the first phase of cluster elongation. The second phase implies that the cluster continues the equilibrium growth in the perpendicular direction (from the pit) until either an equilibrium $s/l$ ratio is established or impingement to other pit happens.

These observations and reasoning maybe correct for the particular case of GS-MBE were propagated to all deposition methods. We have solid objections to it. First of all, only relatively rarefied arrays on very thick hydrogen-rich wetting layers were grown by GS-MBE and investigated in Ref.~\cite{Goldfarb_2006}. Dense arrays obtained by UHV MBE grow in absolutely different manner (Figs.~\ref{fig:array_360C} and \ref{fig:array_530C}). Secondly, as it is seen from Figs.~\ref{fig:array_360C} and \ref{fig:array_530C}, no pits are available in the wetting layer grown by UHV MBE at moderate temperatures.
Thirdly, according to our observations, which are as a rule carried out with atomic resolution, neither pits nor steps are required for cluster nucleation and growth by  UHV MBE (Figs.~\ref{fig:array_360C} and \ref{fig:array_530C}, Ref.~\cite{photon-2008lt}). And at last, the described tricky evolutions of hut would leave their imprints on the cluster shape and structure which would be seen in a good microscope. Unfortunately, this is not the case and no marks of the above evolutions are seen even at atomic resolution.

A model competing with those of equilibrium-driven elongation is usually referred to as  kinetically driven elongation \cite{Island_growth,Kastner,Voigt_Review}. In this model, one of the pyramid facets begins to  grow by chance due to  a fluctuation. Coverage by a monolayer decreases the likelihood of the next attachment of a Ge adatom to newly appeared trapezoid facets  whereas the probability of attachment to triangular ones remains constant. Due to increase of areas of the  trapezoid facets and consequently the barrier of adatom attachment,  further elongation in the randomly chosen $\langle$1\,0\,0$\rangle$ direction goes on with decreasing probability of Ge atom attachment to the trapezoid facets and hence with decreasing rate of the in-height growth in comparison with the rate of the longitudinal one. Unfortunately this model appeared to disagree with experiments. Observations of Goldfarb {\it et al.} \cite{Goldfarb_2006} indicate that in case of interaction of pits and huts  the two  $\langle$1\,0\,0$\rangle$ directions of elongation are not equally probable as it follows from the model of  kinetically driven elongation. Our elaborated observations, which are presented below,  do not support this model either. 

Summarizing this brief historical review we would like to emphasize the following. To date, main milestones on the pathway of the  Stranski-Krastanow growth of Ge film on Si(001) at moderate temperatures ($\apprle$\,550\textcelsius) from the pure silicon surface to hut clusters appearance are recognized as consecutive steps of the strain relief. They are following: $(2\times 1) \rightarrow (2\times n) \rightarrow (M\times N) \rightarrow huts$.  However, despite the efforts made there is no clarity  in the issues of hut nucleation and further transformation. The process of hut nucleation appeals for detailed experimental investigation at different growth conditions by instruments assuring atomic resolution  and using different deposition methods. Formation and longitudinal growth of huts  have not been understood thus far. Elongation of pyramids {\it has never been unambiguously observed in experiments}. The best of the two theoretical models explaining the elongation of wedges also has not been chosen yet. Issues of evolution of the cluster arrays during Ge deposition has been passed over by researchers too. There has been no systematic investigations of the stages of this very important and complicated process presented in literature. A final phase of the array growth cycle---growth at high coverages and transition to the 2D mode---has never been in focus of investigations.
So, we can conclude now that despite the widely adopted standpoint,  investigations on the discussed problem are still far from completion.

The above analysis of the literature puts a number of new questions the most obvious of which are as follows. Firstly, it is unclear whether the structure of pyramids and wedges is identical  and whether they
 belong to the same morphologically uniform class. Secondly,  does the structure of pyramids and wedges coincide at the moment of nucleation or have they  got different structures already at the stage of occurrence and consequently  different nuclei?  Thirdly, are their roles the same in the array formation and evolution? Fourthly, what are the driving forces of their evolution during the array growth? Why the results of the evolution are different for pyramids and wedges? And at last, is it possible to control the cluster evolution to obtain uniform and defectless dense arrays suitable for the industrial applications?
 
In the current paper, we shall study a part of the above issues focusing mainly on the morphology of hut clusters and partly on the growth cycle of the dense arrays.

\section{\label{sec:setup}Experimental}

\begin{figure}[h]
\centering
\includegraphics[scale=1.2]{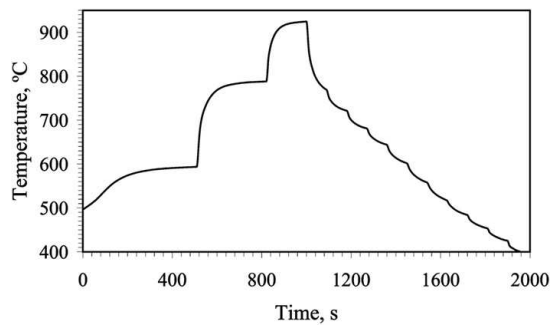}
\caption{\label{fig:cycle} A diagram of the final thermal treatment of the Si substrates during surface deoxidizing in the MBE chamber.}
\end{figure}

The experiments were made using an integrated ultrahigh vacuum system based on the Riber EVA~32 molecular beam epitaxy chamber  coupled through a transfer line with the STM GPI-300 \cite{gpi300} scanning tunnelling microscope (STM). This instrument enables the STM study of samples with atomic resolution at any stage of Si surface cleaning and MBE growth. The samples can be consecutively moved into the STM chamber for analysis and back into the MBE chamber for further processing never leaving the UHV ambient and their surfaces stay atomically clean over whole the experiment.
A procedure of the sample preparation was as follows: initial substrates were 8$\times$8 mm$^2$ squares cut from B-doped CZ Si$(100)$ wafers (\,$p$-type,  $\rho = 12~\Omega\,$cm). After washing and chemical treatment following a standard procedure described elsewhere (see e.g. Refs.~\cite{Report_01-303,etching}) the silicon substrates mounted on the molybdenum STM holder and clamped with the tantalum fasteners were loaded into the  airlock and transferred to the preliminary annealing chamber where outgassed at  the temperature of around $565^\circ$C and the pressure of about  $5\times 10^{-9}$ Torr for about 24 hours. After that the substrates were moved for final treatment into the MBE chamber evacuated down to about $10^{-11}$\,Torr. There were two stages of annealing in the process of substrate heating in the MBE chamber\,--- at $\sim 600^\circ$C for $\sim 5$ min. and at $\sim 800^\circ$C for $\sim 3$ min. (Fig.~\ref{fig:cycle}). The final annealing at the temperature greater than $900^\circ$C was carried out for nearly $ 2.5$ min. with the maximum temperature of about $ 925^\circ$C ($\sim 1.5$~min.). Then the temperature was rapidly lowered to about $ 750^\circ$C. The rate of the further cooling down was around $0.4^\circ$C/s. The pressure in the MBE chamber enhanced to nearly $2\times 10^{-9}$ Torr during the process.

The surfaces of the silicon substrates were completely purified of the oxide film as a result of this treatment. The high-order $(8\times n)$ surface reconstruction described in Ref.~\cite{our_Si(001)_en} was always revealed by the STM on the deoxidized substrates whereas the reflected high energy electron diffraction (RHEED)\footnote{RHEED is usually applied in the MBE vessels for monitoring the surface perfection \cite{Brunner} e.g. during the deoxidizing process.}  patterns obtained from the same purified surfaces always corresponded to either  $(2\times 1)$ or $(4\times 4)$ surface reconstruction \cite{Si(001)_ICDS-25}. This observation is in a good agreement with the model brought forward by us in Ref.~\cite{our_Si(001)_en} as well as with the generally accepted opinion based on the RHEED measurements that the $(2\times 1)$ reconstruction is formed on the Si(001) surface due to deoxidization in an MBE chamber \cite{Smagina}.

Germanium was deposited directly on the purified silicon surface from the source with the electron beam evaporation. The rate of Ge deposition was about $0.15$~\r{A}/s, $h_{\rm Ge}$ was varied from 6~\r{A} to 14~\r{A} for different samples. The deposition rate and the effective Ge film thickness $h_{\rm Ge}$ were measured by the graduated in advance XTC film thickness monitor with the  quartz sensor installed in the MBE chamber. The substrate temperature $T_{\rm gr}$ was $360^\circ$C or $530^\circ$C during the process. The pressure in the MBE chamber did not exceed $10^{-9}$ Torr during Ge deposition. The rate of the sample cooling down to the room temperature was approximately $0.4^\circ$C/s after the deposition. 

The samples were heated by Ta radiators from the rear side in both preliminary annealing and MBE chambers. The temperature was monitored with chromel-allimel and tungsten-rhenium thermocouples in the preliminary annealing and MBE chambers, respectively. The thermocouples were mounted in vacuum near the rear side of the samples and {\it in situ} graduated beforehand against the IMPAC~IS\,12-Si pyrometer which measured the sample temperature through chamber windows with an accuracy of \textpm $(0.3\,\%~T^\circ{\rm C} + 1^\circ$C).

The atmosphere composition in the MBE camber was monitored using the SRS~RGA-200 residual gas analyzer before and during the process.

After Ge deposition and cooling, the prepared samples were moved for analysis into the STM chamber in which the pressure did not exceed $10^{-10}$ Torr. The STM tip was {\it ex situ} made of the tungsten wire and cleaned by ion bombardment \cite{W-tip} in a special UHV chamber connected to the STM chamber. The images were obtained in the constant tunneling current mode at the room temperature. The STM tip was zero-biased while the sample was positively or negatively biased for empty or filled states mapping. 

The WSxM software \cite{WSxM} was used for processing of the STM images.

\begin{figure}[h]
\centering
\includegraphics[scale=1]{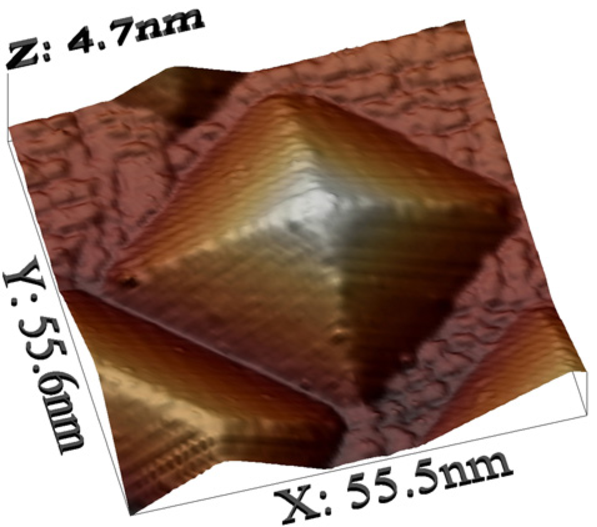} (a) 
\includegraphics[scale=1]{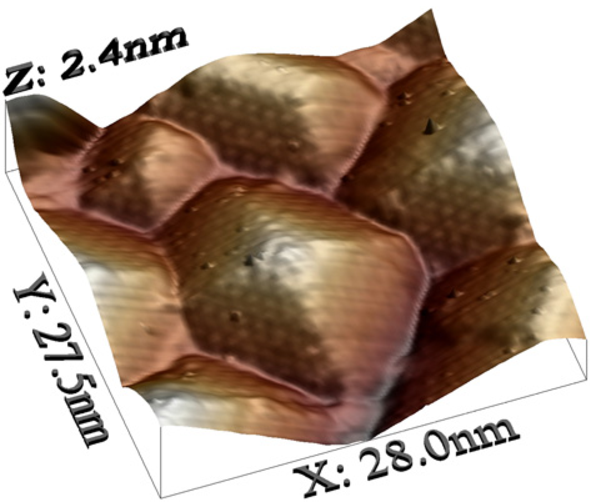} (b)\\
\includegraphics[scale=1]{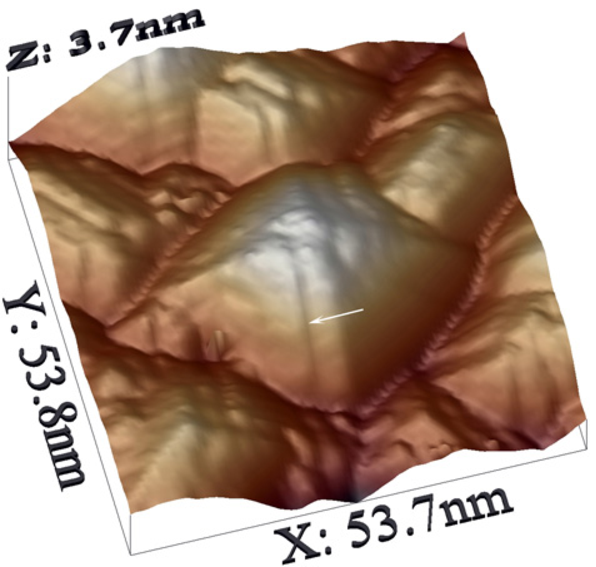} (c)
\includegraphics[scale=1]{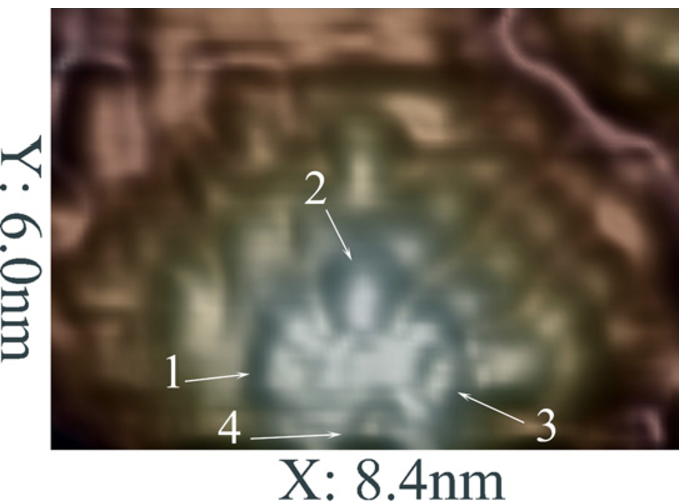} (d)
\includegraphics[scale=1.5]{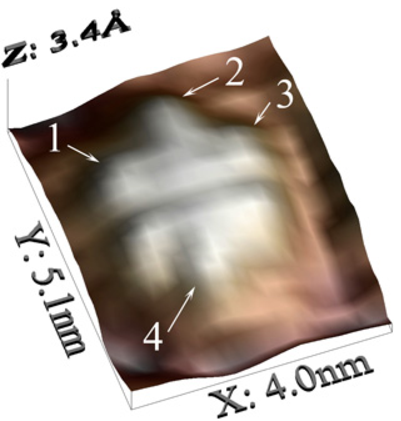} (e)
\caption{\label{fig:pyramid}STM micrographs of the pyramidal Ge clusters; completely shaped clusters: $T_{\rm gr}$ = 530$^\circ$C, $h_{\rm Ge} = 11$~\AA~(a); $T_{\rm gr}$ = 360$^\circ$C, $h_{\rm Ge}$ = 10 \AA~(b); a cluster with unfinished facets (marked by an arrow): $T_{\rm gr}$ = 360$^\circ$C, $h_{\rm Ge} = 14$~\r{A} (c); a structure of vertex, sides and edges (d) and a structure of  a nucleus of 1 monolayer high over the wetting layer  (e): $T_{\rm gr} = 360^\circ$C, $h_{\rm Ge} = 6$~\r{A} (corresponding features on toppings are marked by the same numerals).}
\end{figure}

\section{\label{sec:Class}Classification}

\subsection{\label{sec:main}Main species of the clusters}

\subsubsection{\label{sec:wedge}Pyramidal and wedge-shaped clusters}\

As mention above an array of the self-assembled germanium hut clusters formed on the Si(001) surface consists of a set of morphologically different clusters. All the clusters have the edges of bases oriented along the $\langle 100\rangle$ directions in common. Yet in spite of the apparent variety of the cluster forms (Figs.~\ref{fig:array_360C} and \ref{fig:array_530C}) an analysis of the STM images gives evidence that only two main species of the hut clusters exist---ones having square bases and shapes of the regular pyramids and those with rectangular bases which have shapes of the wedges.
We have already cited the paper by Mo {\it et al.} \cite{Mo}, in which both species of the hut clusters were described for the first time, as well as a number of publications which investigated the details of their formation \cite{Tersoff_Tromp,  Nucleation, Goldfarb_JVST-A, Island_growth, Kastner,  Goldfarb_2005, Goldfarb_2006}. Unfortunately their structure has not been yet clearly visualised and identified.

\begin{figure}[h]
\centering
\includegraphics[scale=1]{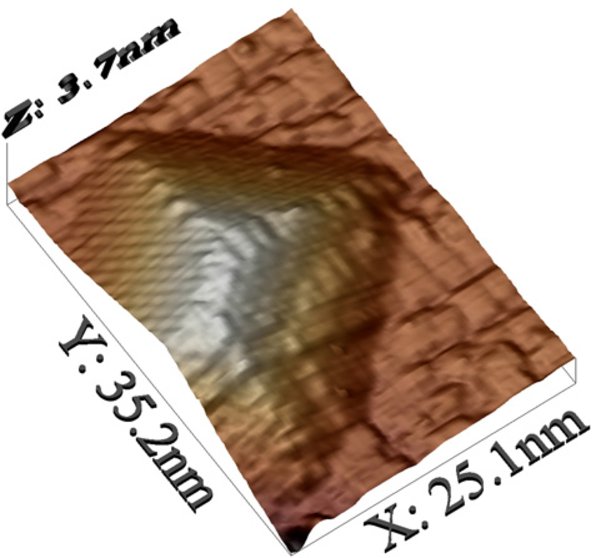} (a)
\includegraphics[scale=1]{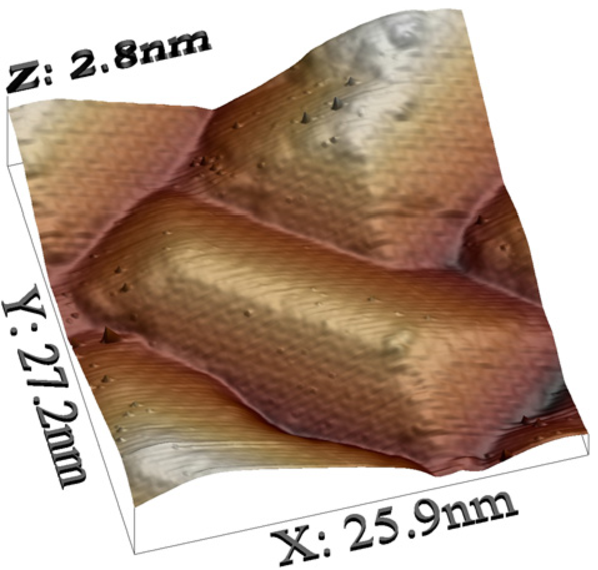} (b)\\
\includegraphics[scale=1]{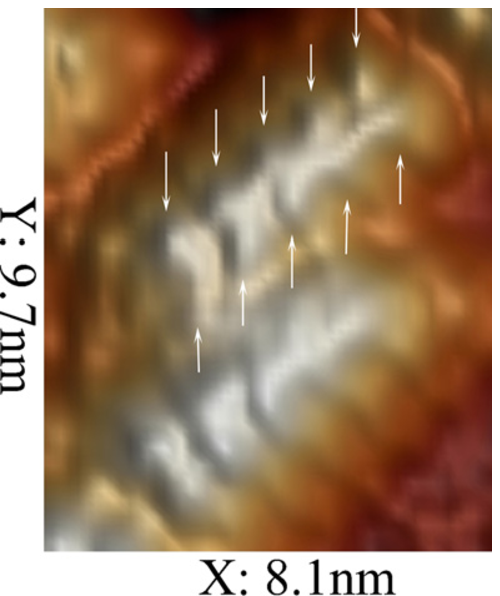} (c)
\caption{\label{fig:wedge}STM images of the wedge-like Ge clusters; a cluster with an unfinished side: $T_{\rm gr}$ = 530$^\circ$C, $h_{\rm Ge} = 11$~\r{A}~(a); the entirely formed cluster: $T_{\rm gr}$ = 360$^\circ$C, $h_{\rm Ge} = 10$~\r{A}~(b); a structure of the ridges (two closely neighbouring clusters), the shifted features are marked by the arrows: $T_{\rm gr} = 360^\circ$C, $h_{\rm Ge} = 8$~\r{A}~(c).}
\end{figure}

Let us dwell on the description of each species of the clusters in more detail.

Typical high resolution STM images of {\em the pyramidal clusters} in the arrays with different growth parameters are shown in Fig.~\ref{fig:pyramid}. A regular shape of the clusters is clearly seen in the pictures (a) and (b) presenting the arrays obtained at $T_{\rm gr} = 530^\circ$C, $h_{\rm Ge} = 11$~\r{A} and $T_{\rm gr} = 360^\circ$C, $h_{\rm Ge} = 10$~\AA, a fine structure of the faces is resolved as well as the $(M \times N)$ patched structure of the wetting layer \cite{Iwawaki_initial,Wu,Voigt,Wetting}. Lines of the solidifying steps are revealed for the first time on the cluster incomplete faces in the image (c) of the array obtained at $T_{\rm gr} = 360^\circ$C, $h_{\rm Ge} = 14$~\r{A} (one of them is marked by arrow in the image). 
Fig.~\ref{fig:pyramid}(d) shows a vertical view of the small pyramid grown at $T_{\rm gr} = 360^\circ$C, $h_{\rm Ge} = 6$~\r{A} and having only 5 monolayers height over the wetting layer. A fine structure of the pyramid vertex and edges as well as the stepped structure of its \{105\} facets are  resolved in detail in the image (d).  And at last, the same structure as that seen in Fig.~\ref{fig:pyramid}(d) on the pyramid vertex is clearly resolved in the image (e) of the pyramid nucleus (1 monolater high over the wetting layer) situated on the block of the Ge $(M \times N)$ surface.\footnote{Note also that similar configuration  of the pyramid  apex   can be discerned by an attentive observer  in  Fig.~\ref{fig:pyramid}(b) which presents an image of a ``ripe''   pyramid in the well developed array.}

Fig.~\ref{fig:wedge}  demonstrates STM images of {\em the wedge-like clusters}. Being the hut clusters they are bounded by \{105\} planes, i.e. their heights are to the base widths as 1:10. A distinctive feature of this species of the clusters is that their base lengths are not connected with cluster heights and are rather random. To some extent, the base lengths of the wedge-like cluster depends on its nearest neighbours. Nevertheless, it is impossible to confidently point out the factors which affect the lengths of the Ge wedges based upon the available data. Reasoning from the results of the STM image analysis it may only be asserted that their base length-to-width ratio is distributed randomly and rather uniformly in the interval from a little greater than 1 to more than 10. 

\begin{figure}[h]
\centering
\includegraphics[scale=0.4]{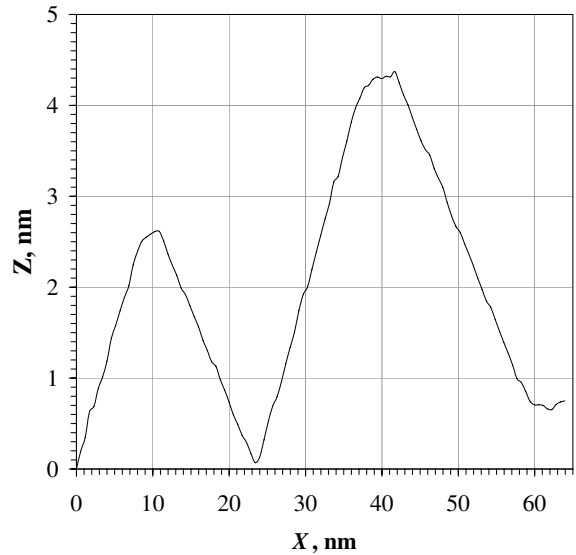} 
\caption{\label{fig:profile1} Profiles of the neighbouring wedge (the left one, taken along the short side of base) and pyramid (the right one) shown in Fig.~\ref{fig:pyramid}(a).}
\end{figure}

As long as the factors determining the base lengths of the wedges are unclear it is unknown if it is possible to control the array growth parameters in such a way to minimize the dispersion of the base lengths of the wedge-like clusters. Moreover, the factors governing the nucleation of either pyramidal  or wedge-like clusters of very different lengths on the wetting layer are also vague. The words about the role of the strain in the wetting layer themselves explain nothing. It stays unclear why the clusters with different shapes (or even symmetry) may arise in a very close vicinity to one another. Perhaps the nucleation process and consequently the strain field (its distribution and local symmetry) are controlled by the underlying Si surface of the Si/Ge interface---its reconstruction before and defect (in particular, vacancy) distribution before and in the process of Ge deposition---especially in the case of the low-temperature processes. This is one of the reasons to consider the surface pre-growth treatment as a key to the controllable Ge/Si heterostructure formation process. That is why the {\it in situ} studies of the Si surface on the atomic scale immediately before the Ge deposition as well as the investigations of its influence on the deposited Ge layer should be considered as a task of high importance. 

Fig.~\ref{fig:wedge}(c) demonstrates the fine structure of ridges of two close wedge-like clusters ($T_{\rm gr} = 360^\circ$C, $h_{\rm Ge} = 8$~\r{A}). It is interesting that one can descry the same configuration of the ridge in the first published image of the hut cluster  (see Fig.~2 in Ref.~\cite{Mo}).

The STM images of fine structure of the vertexes and the ridges,  similar to those shown in Figs.~\ref{fig:pyramid}(d, e) and \ref{fig:wedge}(c), helped us propose structural diagrams of both species of the clusters \cite{photon-2008lt,photon-2008ag}. It is clearly seen from the images that {\it the fine structures of   apexes of the clusters are different}. The features in the uppermost parallel rows on the ridges of the wedge-like clusters  are shifted with respect to one another. (They are marked by the rows of the shifted arrows in the STM image.) We interpret these features as Ge dimer pairs in accordance with the simple structural model of the hut cluster facets (PD model) proposed by Mo {\it et al.} \cite{Mo}. Similar features in the images of vertexes of the pyramids are gathered in the straight rows. The difference of symmetry of  the pyramid vertex from the symmetry of the elementary unit of the wedge ridge is distinctly evident when compared Figs.~\ref{fig:pyramid}(e) and \ref{fig:wedge}(c).

As we have already mentioned above, the difference of the atomic structure causes a ban of shape transitions between the pyramidal and wedge-shaped hut clusters which were intensively discussed in the literature \cite{Kastner,  Goldfarb_2005, Goldfarb_2006}. In addition, particular nuclei should be sought for pyramidal and wedge-shaped clusters. The question appears as well, why two structurally different species of hut clusters arise on the wetting layer.

Fig.~\ref{fig:profile1} shows the cross section profiles of the adjacent wedge-like and pyramidal clusters presented in Fig.~\ref{fig:pyramid}(a). Both clusters are seen to have equal ratio of the base width to the cluster height close to 10.  The base sides of the pyramidal cluster and the base length of the wedge-like one are nearly equal whereas the base width of the wedge-like cluster is by about 1.6 times less than the sides of the pyramid base. It is a common rule which is not affected by the length of a particular wedge-like cluster: The pyramidal clusters are usually higher than the wedge-like ones  if $h_{\rm Ge}$ is high enough (see also, e.g., images (a) to (c) in Fig.~\ref{fig:pyramid}). This observation may be explained if height limitation is assumed to occur for the wedge-like clusters at $h_{\rm Ge}$ greater than some value and not to occur for the pyramids. We do observed such limitation for the wedge-shaped clusters (see below) and  did not succeed to fix the height limitation of the pyramidal ones.\footnote{Perhaps, there is no height limitation for pyramids and namely they give rise to large clusters affecting the properties of the Ge/Si(001) heterostructures and classified by us as one of the types of defects of arrays (see report \cite{Report_01-303} or an article on defects of arrays  \cite{defects_ICDS-25}).}

\begin{figure}[h]
\centering
\includegraphics[scale=0.4]{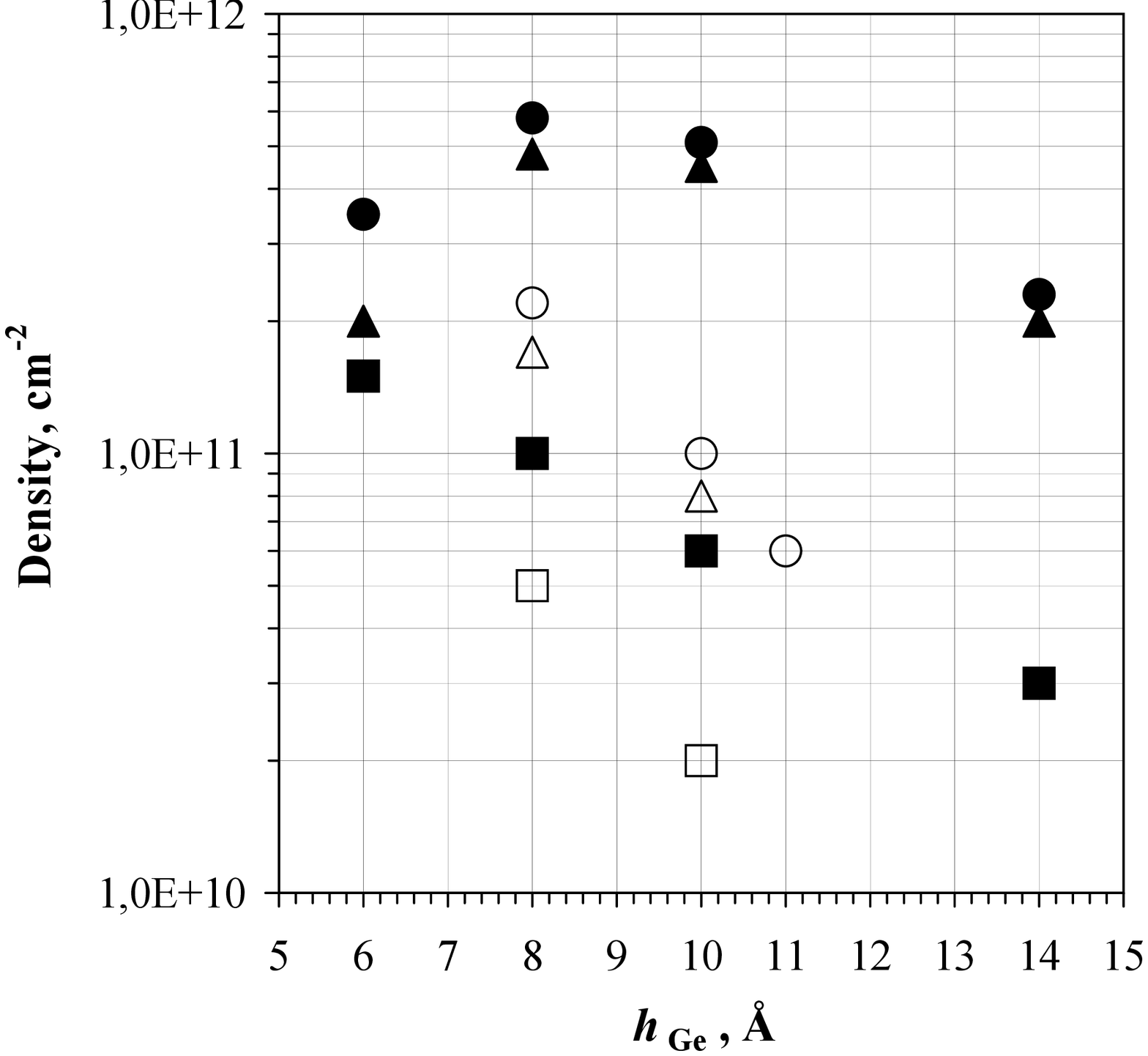}(a)
\includegraphics[scale=0.44]{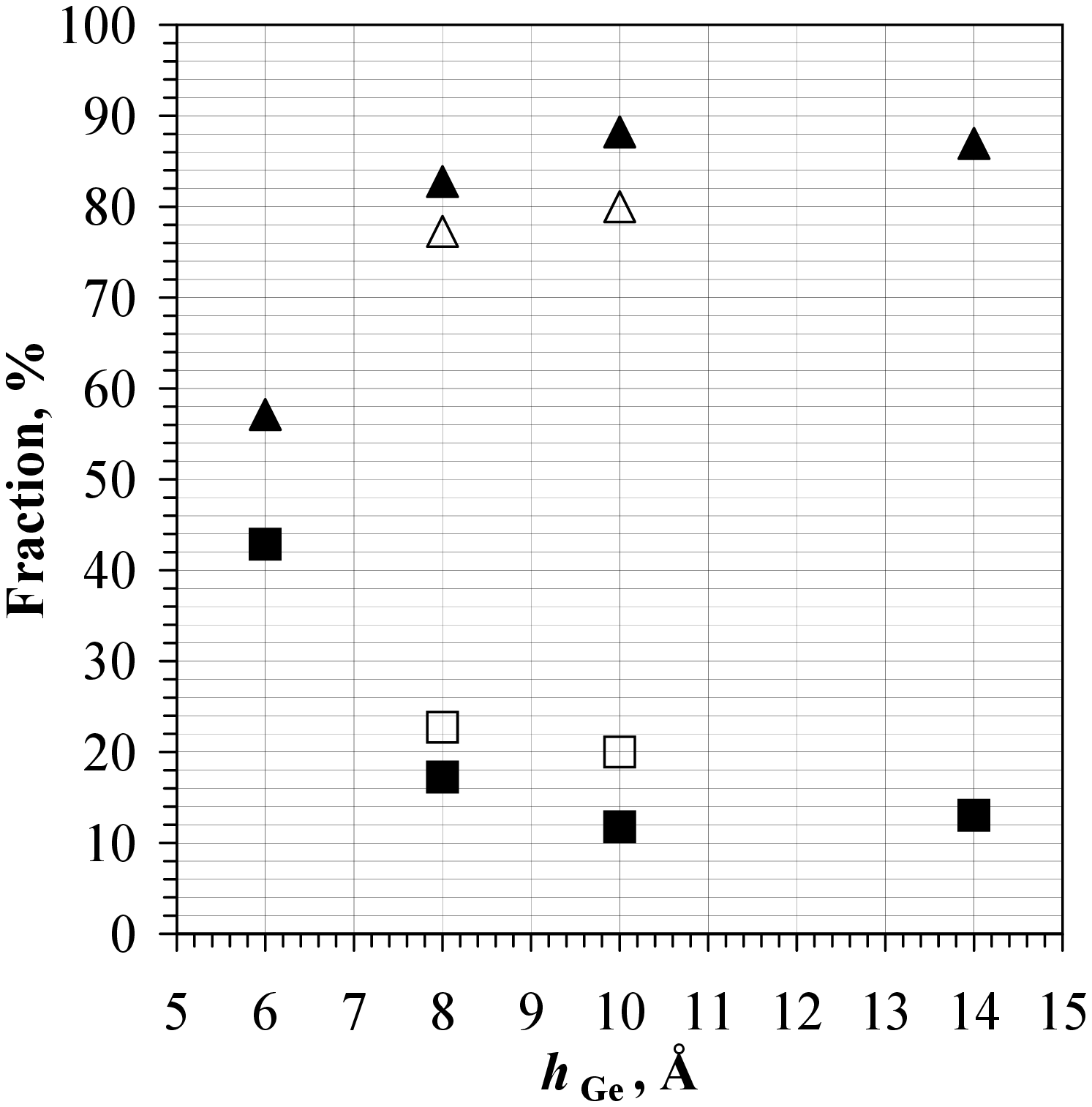}(b)
\caption{\label{fig:density} The density  of the Ge clusters in the arrays~(a) formed at $T_{\rm gr} = 530^\circ$C (designations: $\square$~marks the pyramids, $\triangle$~corresponds to the wedges, $\Circle$~means the total value) and $T_{\rm gr} = 360^\circ$C ($\blacksquare$~designates the pyramids, $\blacktriangle$~means the wedges, $\CIRCLE$~is the total density). The fraction  of the pyramidal and wedge-like Ge clusters in the arrays~(b), $T_{\rm gr} = 530^\circ$C  ( $\square$~marks the pyramids, $\triangle$~designates the wedges), $T_{\rm gr} = 360^\circ$C ($\blacksquare$~corresponds to the pyramids, $\blacktriangle$~designates the wedges). Both graphs are plotted {\it vs} $h_{\rm Ge}$.}

\end{figure}

\subsubsection{\label{sec:fraction}Density and fraction of each species of clusters in the arrays}\

It was observed  that the wedge-like and pyramidal clusters are  different not only in their atomic structure and geometrical shapes. The wedge-like clusters dominate in the arrays formed at low temperatures, and their fraction grows with the growth of $h_{\rm Ge}$ (Fig.~\ref{fig:density}).

Fig.~\ref{fig:density}(a) plots the dependence of the cluster density on $h_{\rm Ge}$ for different clusters in the arrays grown at  $360^\circ$C and $530^\circ$C. It is seen that for $T_{\rm gr} = 360^\circ$C the density of wedges rises starting from $D_{\rm w}\approx 1,8\times 10^{11}$~cm$^{-2}$ at the beginning of the three-dimensional growth of Ge (the estimate is obtained by data extrapolation to $h_{\rm Ge}= 5$~\r{A}) and  reaches the maximum of $\sim 5\times 10^{11}$~cm$^{-2}$ at $h_{\rm Ge} \sim 8$~\r{A}, the total  density of clusters at this point $D_{\rm \Sigma} \sim 6\times 10^{11}$~cm$^{-2}$ is also maximum. Then both  $D_{\rm w}$ and $D_{\rm \Sigma}$ slowly go down until the two-dimensional growth of Ge starts at $h_{\rm Ge} \sim 14$~\r{A} and  $D_{\rm \Sigma}\approx D_{\rm w} \sim 2\times 10^{11}$~cm$^{-2}$ (the contribution of pyramids $D_{\rm p}$ to  $D_{\rm \Sigma}$ becomes negligible---~$\sim 3\times 10^{10}$~cm$^{-2}$---at this value of $h_{\rm Ge}$). 
 The pyramid density exponentially drops for $T_{\rm gr} = 360^\circ$C as the value of $h_{\rm Ge}$ grows ($D_{\rm p} \approx 5\times 10^{11} \exp\{-2.0\times 10^7\,h_{\rm Ge}\}$, $h_{\rm Ge}$ is measured in centimeters). The maximum value of $D_{\rm p} \approx 1.8\times 10^{11}$~cm$^{-2}$ obtained from extrapolation to $h_{\rm Ge} = 5$~\r{A} coincides with the estimated initial value of $D_{\rm w}$.

For $T_{\rm gr} = 530^\circ$C, the total density of clusters exhibits the same trend as $D_{\rm p}$ for $T_{\rm gr} = 360^\circ$C, $D_{\rm \Sigma} \approx 7\times 10^{12} \exp\{-4.3\times 10^7\,h_{\rm Ge}\,[\rm cm]\}$. The maximum (initial) value of $D_{\rm \Sigma}$ is estimated as $8\times 10^{11}$~cm$^{-2}$ by extrapolation to $h_{\rm Ge} = 5$~\r{A}.

The graphs of cluster fractions in the arrays versus $h_{\rm Ge}$ are presented in Fig.~\ref{fig:density}(b). For $T_{\rm gr} = 360^\circ$C, portions of pyramids and wedges initially very close ($\sim 50\%$ at $h_{\rm Ge}\sim 5$~\AA) rapidly become different as $h_{\rm Ge}$ rises. The content of pyramids monotonically falls. The fraction of the wedge-like clusters is approximately $57\%$ at the early stage of the array growth ($h_{\rm Ge} = 6$~\r{A}) and becomes $82\%$ at $h_{\rm Ge} = 8$~\r{A}. At further growth of the array, the content of the wedges reaches the saturation at the level of approximately $88\%$ at $h_{\rm Ge} = 10$~\r{A}. 

At moderate values of $h_{\rm Ge}$, the proportion of pyramids to wedges for $T_{\rm gr} = 530^\circ$C was found to be nearly the same as for $T_{\rm gr} = 360^\circ$C.  The content of the pyramidal clusters in the array is about $20\%$ at $h_{\rm Ge} = 8$ and 10~\r{A}. 

The inference may be made from this observation that contrary to the intuitively expected from the consideration of symmetry, the wedge-like shape of the clusters is energetically more advantageous than the pyramidal one, and the more advantageous the more Ge atoms (and the more the number of atomic layers) constitute the cluster. 
The probability of nucleation appears to be close to 1/2 for both wedges-like and pyramidal clusters at the initial stage of the array formation and low growth temperatures. Then, as the array grows, the formation of pyramids becomes hardly probable and most of them, which have  already been formed, vanish whereas the nucleation and further growth of wedges continues (Fig.~\ref{fig:array_360C}). The Ge pyramides on the Si(001) surface appear to be less stable species than the wedges and in accordance with the ``bourgeois principle'' (``the survival of the fittest'') they loose their substance in favour of the wedge-like clusters. 

At higher temperatures, no nucleation of new clusters was observed in the process of the array growth (Fig.~\ref{fig:array_530C}). The  ``bourgeois principle'' decreases the cluster density in the arrays and increases their sizes despite the species they belong to. 

It should be noted that the above analysis demonstrates that pyramidal and wedge-like clusters are really different objects which belong to different cluster species rather than the varieties of the same structurally uniform species as it is usually postulated in the literature \cite{Mo,Island_growth, Kastner,Goldfarb_2005,Goldfarb_2006}.

Remark also that at $T_{\rm gr} = 360^\circ$C and the flux of Ge atoms ${\rm d} h_{\rm Ge}/{\rm d}t = 0.15$~\r{A}/s, the point $h_{\rm Ge} = 10$~\r{A} is particular. Not only the fraction of pyramids saturates at this point but the array in whole has the most uniform sizes of the clusters composing it (Fig.~\ref{fig:array_360C}). This is concluded by us not only on the basis of analysis of the STM images of the Ge/Si(001) arrays but also from the data of the Raman light scattering by the Ge/Si heterostructures with different low-temperature arrays of Ge quantum dots \cite{our_Raman_en,Raman_conf}. We refer to such arrays as optimal.

A qualitative model accounting for the presence of the particular point at the low-temperature array growth is simple. The case is that at low enough temperatures of the array growth, the new Ge cluster nucleation competes with the process of growth of earlier formed clusters. The height of the clusters (at least the dominating wedge-like ones) is observed to be limited by some value governed by $T_{\rm gr}$. At small $h_{\rm Ge}$, Ge clusters are small enough and the distances between them are large enough compared to the Ge atom (or dimer) diffusion (migration) length on the surface for nucleation of new clusters on the Ge wetting layer in the space between the clusters (Fig.~\ref{fig:array_360C}(a,\,b)). At $h_{\rm Ge} = 10$~\r{A} and the above ${\rm d} h_{\rm Ge}/{\rm d}t$  value, the equilibrium of parameters (cluster sizes and distances between them, diffusion length at given temperature, Ge deposition rate, etc.) sets in, the rate of new cluster nucleation is decreased and the abundant Ge atoms are mainly spent to the growth of the available clusters (Fig.~\ref{fig:array_360C}(c)). After the clusters reach their height limit and in spite of it, Ge atoms continue to form up their facets. As soon as most of the clusters reach the height limit, nucleation of new clusters becomes energetically advantageous again and the nucleation rate rises. A second phase of clusters appears on the wetting layer and fills whole its free surface as $h_{\rm Ge}$ is increased (Fig.~\ref{fig:array_360C}(d)). Further increase of $h_{\rm Ge}$ results in two-dimensional growth mode. It is clear now why the array is the most homogeneous (optimal) at $T_{\rm gr} = 360^\circ$C and $h_{\rm Ge} = 10$~\r{A} whereas the dispersion of the cluster sizes is increased at higher and lower values of $h_{\rm Ge}$ because of the small clusters included in the array. It is clear also that the optimal array will appear at different value of $h_{\rm Ge}$ when $T_{\rm gr}$ or ${\rm d} h_{\rm Ge}/{\rm d}t$ are different.

\begin{figure}[h]
\centering
\includegraphics[scale=1]{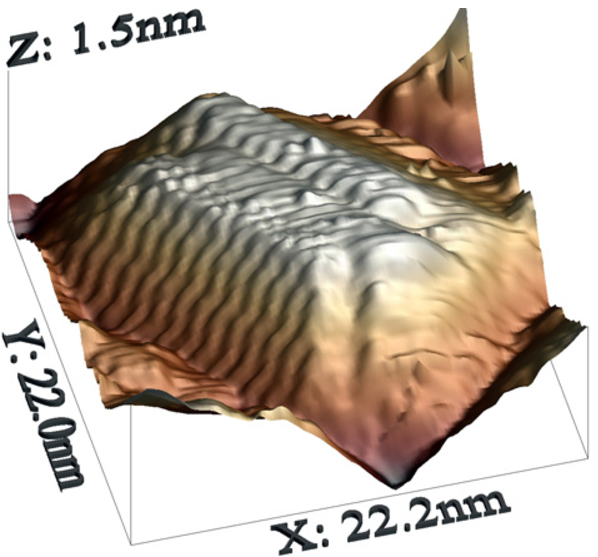} (a)
\includegraphics[scale=1]{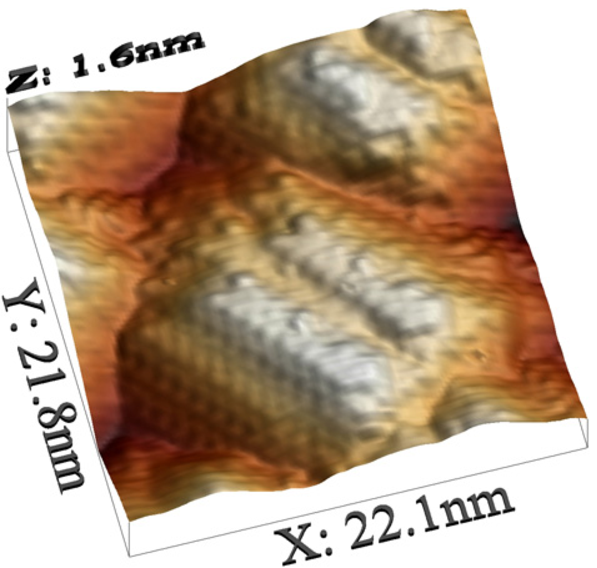} (b)
\caption{\label{fig:obelisk} STM micrographs of the Ge wedges with two ridges (obelisks), a general view of the cluster (a) and a top view of the ridges (b); $T_{\rm gr} = 360^\circ$C, $h_{\rm Ge} = 14$~\AA.}
\end{figure}

\begin{figure}[h]
\centering
\includegraphics[scale=0.5]{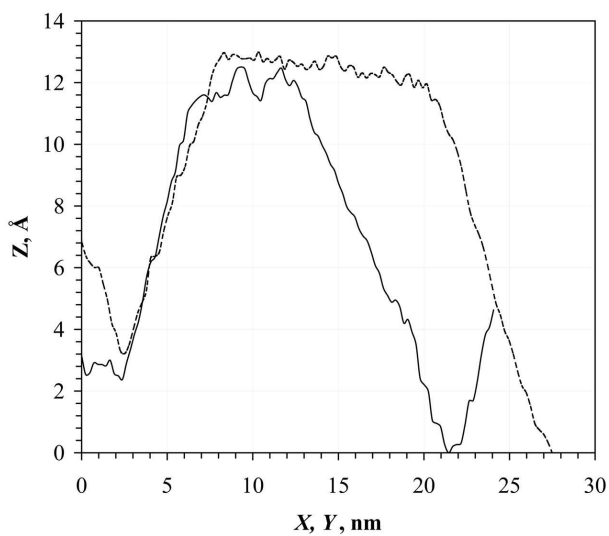}
\caption{\label{fig:profile2} Profiles of the Ge obelisk presented in Fig.~\ref{fig:obelisk}(a) measured across the ridges along the base width (the solid line) and over the left ridge along the base length (the dashed line). Here, as distinct from Fig.~\ref{fig:obelisk}, $X$ and $Y$ mean respectively axes directed along the cluster base width or along its length. Height  is as usually counted from the wetting layer level ($Z=0$). }
\end{figure}

In addition, some threshold value of $T_{\rm gr}$  must exist beyond which the cluster growth process always dominates and the nucleation of clusters happened once will never be repeated. An example of such arrays formed at $T_{\rm gr}  = 530^\circ$Ñ exceeding the threshold value is given in Fig.~\ref{fig:array_530C}.

\subsection{\label{sec:derivative}Derivative species of clusters}

\subsubsection{\label{sec:obelisks} Obelisks (truncated wedges with two ridges)}\

Except for the described above main species of Ge hut clusters, different clusters are also formed on the Si(001) surface which cannot be classified as independent species because they originate from the wedge-like clusters but have specific shapes, particular formation mechanisms, maybe peculiar properties and hence should be marked out in a separate but derivative species. Fig.~\ref{fig:obelisk} shows clusters related to one of the species of derivative clusters---truncated wedge-like clusters with two ridges or obelisk-shaped clusters ($T_{\rm gr} = 360^\circ$C, $h_{\rm Ge} = 14$~\r{A}).

Profiles of the Ge obelisk shown in Fig.~\ref{fig:obelisk}(a) taken along the short and long base sides are presented in Fig.~\ref{fig:profile2}. Although these clusters are the huts and have a slope of the facets $\sim 11.3^{\circ}$ the ratio of the cluster height to its base width is $\sim 0.06$.

\begin{figure}[h]
\centering
\includegraphics[scale=1]{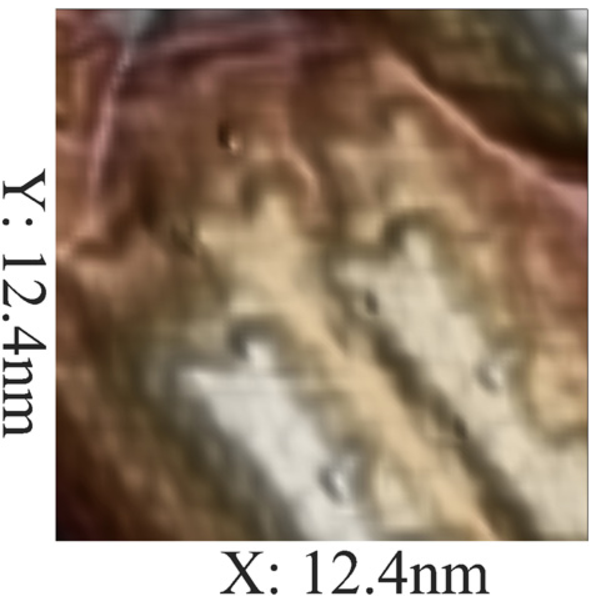} (a)
\includegraphics[scale=1]{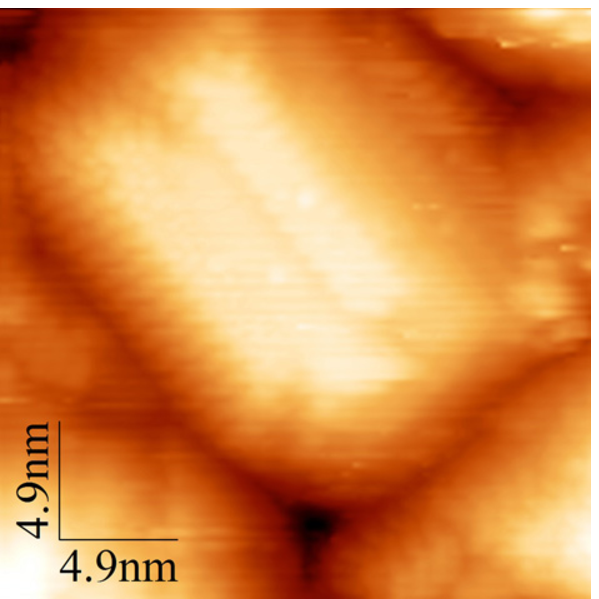} (b)\\
\includegraphics[scale=1.2]{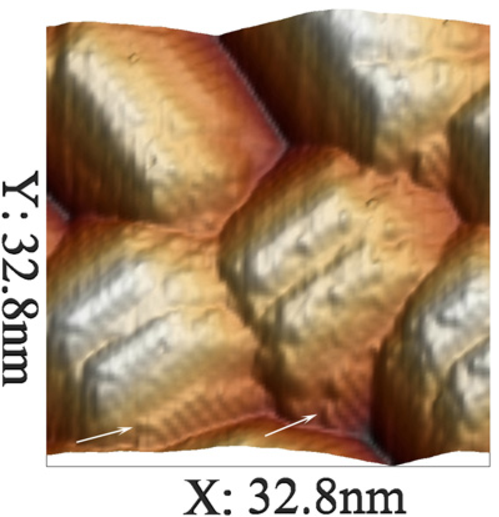} (c)
\includegraphics[scale=1.2]{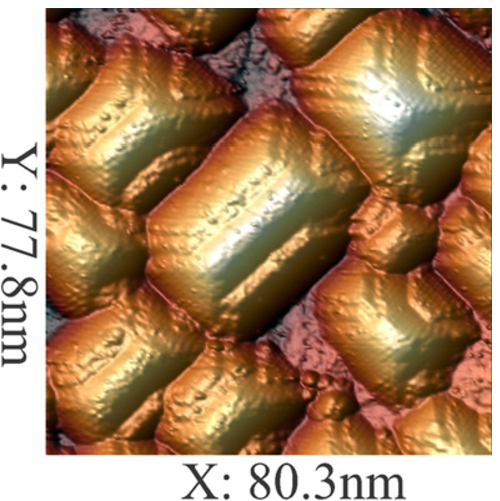} (d)\\
\includegraphics[scale=1]{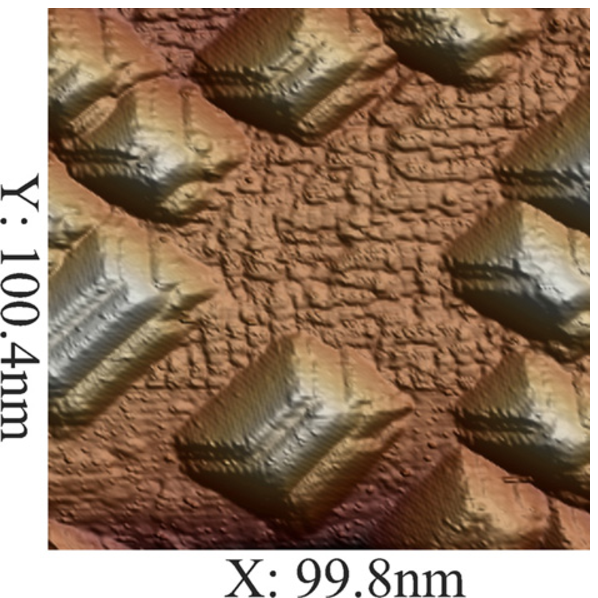} (e)
\includegraphics[scale=1]{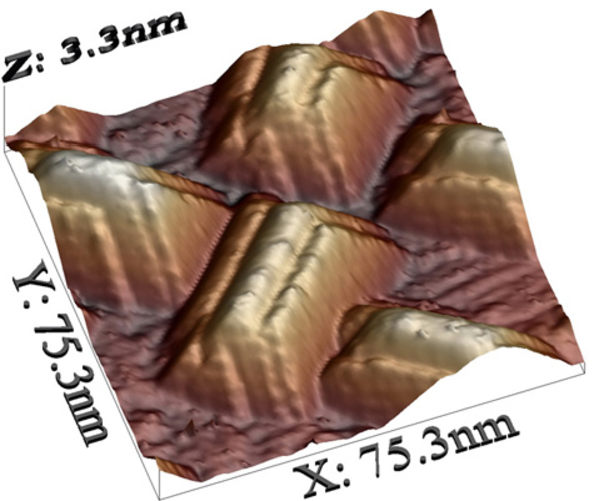} (f)
\caption{\label{fig:top}STM images of the Ge obelisks illustrating the mechanism of their formation; $T_{\rm gr} = 360^\circ$C, $h_{\rm Ge} = 14$~\AA~(a, b, c); $T_{\rm gr} = 530^\circ$C, $h_{\rm Ge} = 8$~\AA~(d) and 
 10~\AA~(e, f).}
\end{figure}

Fig.~\ref{fig:top} presents the STM images of Ge wedges with two ridges explaining the way of their formation. The double ridges are seen to arise as a result of building of the trapezoid facets in the process of the cluster growth. Initially the clusters had the only ridge being ordinary wedge-like ones.  Wedges with double ridges arise because  a limit value of the cluster height exists for any temperature $T_{\rm gr}$. If the limit is reached the further growth of the cluster always goes on by building its trapezoid facets and increasing its width.

This  species of clusters dominates in the arrays at high values of $h_{\rm Ge}$ which depend on the value of $T_{\rm gr}$ (see Figs.~\ref{fig:array_360C} and~\ref{fig:array_530C}).

A unique illustration of the process of the trapezoid facet growth is shown in Fig.~\ref{fig:top}(c). Several (from four to six)  incomplete (001) terraces are seen near the bottoms of the clusters (the dimer pairs  are distinctly resolved, the arrows show these new growing facets in the STM image). These incomplete faces are seen to repeat the shapes of the former faces on which they grow and which are also incomplete. The highest terraces start to grow before the lower ones finish the completion process, as well as  new facets nucleate before the old ones finish the growth. It is also observed  that  the new facets nucleate far from the corners adjoining the base sides. As a result, a compound structure of the trapezoid facets is formed which can be seen in the truncated wedges represented, e.g., in Fig.~\ref{fig:pyramid}(c). We propose the reader to compare the above description with the speculations by Jesson  {\it et al.} which draw a different picture of the facet growth \cite{Island_growth}. First of all, they consider a triangular face as a preferential site of a new facet nucleation explaining in such a way elongation of clusters. Then, according to their model, the facets nucleate in the corners rather than somewhere else, etc. 

We would like to remark that we have not succeeded to observe the growth of the triangular faces at the  temperatures as low as $360^\circ$C.
Nevertheless we observed this process at $T_{\rm gr} = 530^\circ$C. Figs.~\ref{fig:top}(d--f) demonstrate this phenomenon. The peculiarities of the process are as follows: A new facet formation takes place when the cluster has already reached its limit height and its additional  trapezoid facets are well developed. The growing triangular facets are clearly observed on only  one side of the clusters. The triangular faces can nucleate both far from the bottom corner (Figs.~\ref{fig:top}(d)) and close to the corner (Figs.~\ref{fig:top}(e, f)). The growing (incomplete) faces replicate the shape of initial facet even if the latter is complex (composed by intersecting triangles  due to the developed additional trapezoid faces, see Figs.~\ref{fig:top}(e, f) in which the growing faces on the short sides of the truncated wedges are shaped by two combined triangles). It can be deduced from these observations that the described process of formation of new triangular facets  is different from that resulting in the discussed above significant elongation of the wedge-shaped clusters at earlier stages of the cluster growth.

Now we would like to attract the reader's attention to the observed in Figs.~\ref{fig:top}(c,\,e,\,f) formation of the so called ``square based clusters'' from the wedge-like ones which is caused by the extensive growth on the trapezoid facets and the cluster widening. In Fig.~\ref{fig:top}(c), the nearly ``square based clusters'' are seen to be formed because of the widening of the wedges with two ridges. They resemble truncated pyramids but actually preserve the structure of the wedge. The nearly ``square based clusters'' are also seen in the upper left corner of Fig.~\ref{fig:top}(e) and upper right corner of Fig.~\ref{fig:top}(f). These clusters are formed of the formerly wedge-like ones by successive addition of new incomplete facets. Their faces are complex and their shapes are far from the shape of an ideal regular pyramid. Certainly, their structure stays that of the wedge.
The genuine pyramidal cluster revealed in the upper right corner of Fig.~\ref{fig:top}(d) grows nearly uniformly on all four its triangular faces (compare also with Fig.~\ref{fig:pyramid}(c)). 
We would like also to indicate the formation of serial incomplete faces  resolved on the sides of both pyramidal and wedge-shaped clusters presented in Fig.~\ref{fig:top}(d) (even the dimer pairs on the ``parallel steps'' are seen). This process do transform the shape of the clusters and may create nearly ``square based clusters'' from the wedges as well as so called ``rectangular based clusters'' from the pyramids (the latter process may be fancied, e.g., if a pyramid is closely surrounded by its neighbours from all sides except for one and have some room for elongation only in one direction). Of course,  such  transformed clusters are always ``truncated'', have a complex ``stepped'' structure of successive incomplete facets and  atomic structure of apexes characteristic for their precursors, as it is seen in the presented STM images.

\subsubsection{\label{sec:merged}Accreted wedges}\

\begin{figure}[h]
\centering
\includegraphics[scale=1]{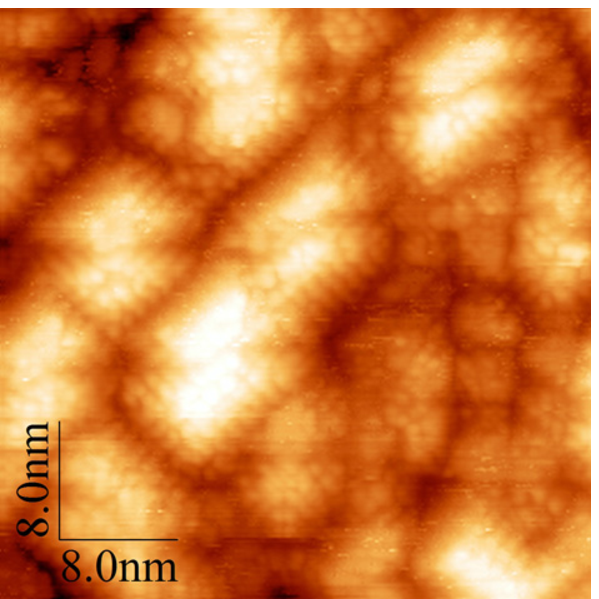} (a)
\includegraphics[scale=1]{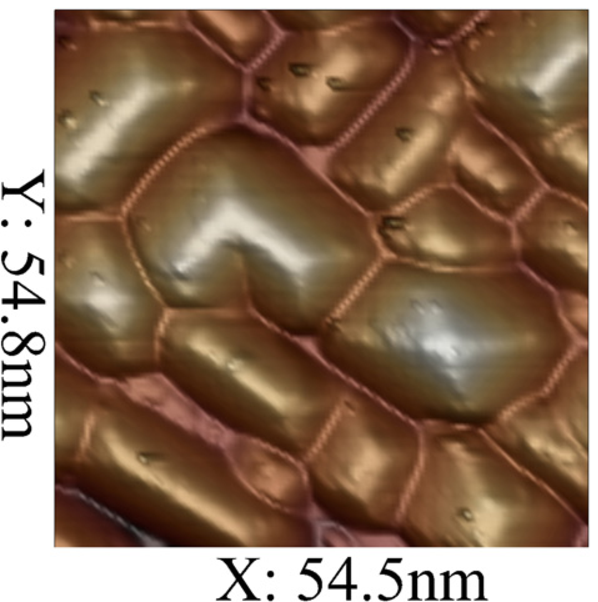} (b)\\
\includegraphics[scale=1]{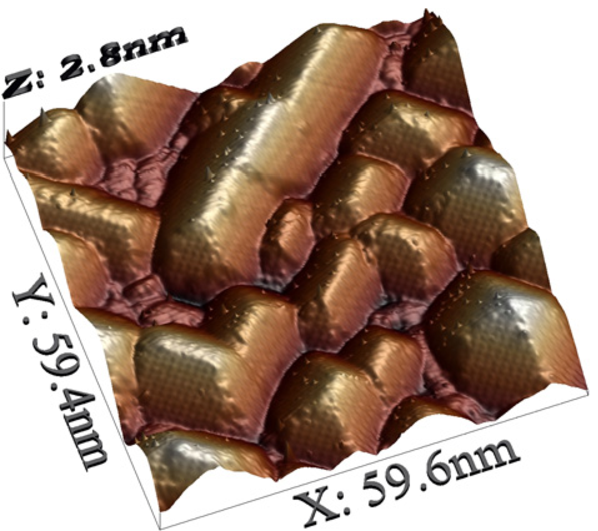} (c)
\caption{\label{fig:accreted}STM images of the accreted together Ge wedges; the accreted clusters borne the clusters with two ridges,  $T_{\rm gr} = 360^\circ$C, $h_{\rm Ge} = 8$~\AA~(a); the accreted clusters given rise to the $\Gamma$-shaped cluster,  $T_{\rm gr} = 360^\circ$C, $h_{\rm Ge} = 10$~\AA~(b); the merged at about half-length wedges formed an extended wedge-like cluster with zigzag on the ridge and the sides,  $T_{\rm gr} = 360^\circ$C, $h_{\rm Ge} = 10$~\AA~(c).}
\end{figure}

In Fig.~\ref{fig:accreted}, the STM images of accreted together wedge-like Ge clusters are shown---accreted clusters which formed ones with two ridges ($T_{\rm gr} = 360^\circ$C, $h_{\rm Ge} = 8$~\AA), accreted clusters which gave rise a $\Gamma$-like one ($T_{\rm gr} = 360^\circ$C, $h_{\rm Ge} = 10$~\AA), clusters accreted approximately at half width formed an extended one with zigzag on the ridge and the facets ($T_{\rm gr} = 360^\circ$C, $h_{\rm Ge} = 10$~\AA) are depicted. 

Like the obelisks the accreted clusters cannot be classified as independent species because they also originate from the wedge-like clusters. However, they also have specific shapes and probably peculiar properties and consequently like obelisks should be separated in a special but derivative species. It was found from the analysis of the STM images obtained at different stages of the array formation that the nuclei of such clusters are situated at the distance of   $\apprle 32$~\r{A} from one another at the initial stage of the array formation in the applied array growth conditions.

From physical viewpoint, these clusters are obviously single and integral objects, the properties of which may be different from the properties of the usual wedge-like clusters. Their influence upon the properties of the arrays in whole is in prospect of further investigations.

The coalescence of clusters is also seen in  Fig.~\ref{fig:pyramid}(c). The truncated wedges are merged or even completely absorbed by the growing pyramid (the pyramids are always greater than the wedges). Such process usually takes place at high values of  $h_{\rm Ge}$ just before the beginning of the two-dimensional growth.

\section{\label{sec:conclusion}Conclusion}

Summarising the above we would like to emphasise the central ideas of the paper.

Morphological investigations and classification of Ge hut clusters forming the arrays of quantum dots on the Si(001) surface at low temperatures in the process of the ultrahigh vacuum molecular beam epitaxy have been carried out using {\it in situ} scanning tunnelling microscopy. 
The study reported in the paper was made in view of the necessity to controllably produce highly uniform and very dense arrays of Ge quantum dots at low temperatures in the process compatible with the CMOS one. Although this task is still far from the solution an important step is made in understanding the object properties to be controlled.

The Ge$/$Si(001) system appeared to be much more sophisticated than it seemed to most of the researchers, and the knowledge about it which is present in the literature now seems to be very deficient and sometimes incorrect. This seems to be the main cause of failure for the last two decades to develop electronic or  photonic devices on the basis of ensembles of Ge quantum dots on the Si(001) surface.

Analysis of the high quality STM images which can be obtained only using an integrated high resolution UHV STM--MBE instrument allowed us to introduce a new classification of germanium hut clusters formed on the Si(001) surface.  The hut clusters were found to be subdivided into four species, two of which are basic and structurally different---the wedge-like and pyramidal clusters---and the rest are derivative---the obelisk-shaped and accreted wedge-shaped clusters. The conclusion was made  that shape transitions between pyramids and wedges are prohibited. The nucleation likelihoods of pyramids and wedges appeared to equal $1/2$ at the initial stage of the array formation. The wedge-like clusters were observed to quickly become the dominating species in the arrays while the pyramidal clusters were found to exponentially rapidly disappear as the arrays grow. 

The derivative types of the clusters have been found to start dominating  at high Ge coverages. The obelisks originate from the wedges as a result of their height limitation and further growth of trapezoid facets. The apexes of the obelisks are formed by  sets of the parallel (001) ridges. 

At low growth temperatures ($360^\circ$C) nucleation of new clusters is observed during the array growth at all values of Ge coverage except for a particular point at which the arrays are  more uniform than at higher or lower coverages. At higher  temperatures ($530^\circ$C) cluster nucleation has not been observed after the initial stage of the array formation.

The growing  trapezoid and triangular cluster facets were visualised. The peculiarities of the facet completion were described. It was shown that the growth of incomplete facets results in a complex structure of the growing hut clusters. 

It was shown also that the uniformity of arrays is governed by the lengths of the wedge-like clusters. This parameter is hardly controllable as distinct to the cluster width which is bound to the cluster height and hence is much more predictable. The cluster lengths are now absolutely unpredictable. Moreover the origin which determines their values is unknown at present. This difficulty requires extensive investigations and intensive efforts to be overcome. But it worth while doing because both stochastic (disordered) and the most promising artificially ordered or self-arranging dense arrays of self-assembled Ge clusters on the Si(001) surface \cite{Ge_QD_crystal} are equally required to be uniform and equally subjected to effect of the Ge wedge length unpredictability.

\ack
The authors appreciate the Science and Innovations Agency of the Russian Federation for funding this research under the State Contract No.\,02.513.11.3130.

\section*{References}

\bibliography{Classification}

\end{document}